# Near-zero Temperature Coefficient of Resistance for a Single-Walled Carbon Nanotube Polymer Nanocomposite

M. Jafarypouria[*], S.G. Abaimov

*Skolkovo Institute of Science and Technology*, *Moscow*, *Russia*

[*]Corresponding author: Milad.Jafarypouria@skoltech.ru

**Abstract**

Temperature dependence of electrical resistance of single-walled carbon nanotube (SWCNTs)/epoxy nanocomposites, which is characterized by a temperature coefficient of resistance (TCR), is experimentally investigated. In the existing literature, there are biased TCR values with non-monotonic temperature dependence, including both negative and positive values. In this study, we demonstrate that the influence of environment temperature can be greatly minimized by ensuring the CNT/epoxy nanocomposite is fully cured along with controlling CNT concentration and selecting an appropriate polymer matrix. In addition, agglomeration of CNTs significantly influences the performance of CNT/polymer nanocomposites. Resistance values remained nearly constant at a CNT concentration of approximately 1 wt.% below the glass transition temperature. It is also observed that in a fully cured CNT/epoxy nanocomposite with the microscale conductive pathway as the main conductance mechanism, the coefficient of thermal expansion (CTE) of the matrix is a critical characteristic, resulting in a consistent increasing trend in TCR.

## 1. Introduction

The integration of carbon nanotubes (CNTs) into polymer nanocomposites offers promising advancements in the development of next-generation nanoelectromechanical sensors, characterized by high precision and sensitivity, lightweight, highly durable and easily adaptable to fit specific shapes and applications [1–15]. For example, sensors made from CNT/polymer nanocomposites can be seamlessly embedded within structural composites for real-time structural health assessment [8]. Additionally, they can be used as highly functional electronic skin (e-skin), capable of simultaneously detecting various stimuli such as temperature and strain [9]. These sensors can also be used for monitoring the degree of cure of polymer during post-curing phase [1], or for real-time monitoring of the manufacturing process and evaluation of the curing degree of polymer composites [11], or in situ resin phase changes and the gel point during the manufacturing process [12], etc. Nonetheless, the conductivity characteristics of CNT/epoxy nanocomposites are influenced by environmental thermal conditions [16,17]. This effect is typically associated with the temperature coefficient of resistance (TCR) [1]:

$$TCR = \frac{1}{R}\frac{dR}{dT} \qquad (1)$$

where *R* is the electrical resistance (or resistivity) of the material and *dR* is its change with the change in temperature *dT*. In the following, we will underline the key factors influencing conductivity, and subsequently the TCR behavior of a CNT/polymer nanocomposite, as identified and studied through experimental research and simulations in the existing literature.

*Morphological structures of nanotubes and tube-based networks*

The electrical properties of a CNT/polymer nanocomposite depend extremely on the morphology of CNTs [18,19] and the morphology of the CNT percolation networks [20–22]. Concurrent methods of modelling allow to build representative volume elements of computationally efficient size due to implementation of periodical boundary conditions and embedded elements [19,20,23]. Contrary to the opinion that embedded elements are available for mechanical analysis only, with appropriate modifications embedded elements can be applied in thermal and electrical analysis as well [24], leading to computationally efficient modelling of many-particle nanocomposites.

The numerical survey carried out by Wang et al. [25] illustrated that the morphology (orientation and diameter) of CNTs is a crucial factor influencing the variation in average junction gaps and piezoresistivity. Gong et al. [26–28] presented a new CNT percolation network model that takes into account the structural distortion of CNT walls at intersecting junctions to explore the effect of CNT morphology on electrical conductivity and piezoresistivity. They demonstrated that the bending of CNTs changes the morphology as well as the effective electrical resistance of CNT percolation networks.

*CNT intrinsic conductance and the tunneling conductance between CNTs*

The inherent conductance of CNTs and the tunneling occurring between them are the two primary forms of conductance in CNT/polymer nanocomposite [15,26–28]. Lomov et al. [29] conducted a numerical modeling study to examine the comparative effects of key phenomena within the temperature range of 300–400 K for both dry and impregnated aligned CNT films. Their analysis revealed that the TCR of the CNTs themselves is the key factor affecting network's TCR, while the TCR of the tunneling contacts plays a secondary role. The dominant influence of the intrinsic conductivity TCR on CNT forest's overall TCR is attributed to the long inter-contact segments of CNTs within the aligned CNT network, which determine the homogenized conductivity of the film. Conversely, experimental findings suggest that interactions between tubes or tubes and the matrix can lead to structural distortions in CNTs [30–33]. These distortions can significantly influence the local electrical configuration, functioning as a solid scattering mechanism that decreases intrinsic conductance. Conversely, they may also lead to a substantial overlap of wave functions, which boosts electron tunneling and improves contact conductance [33]. Therefore, neglecting the structural distortions of CNTs could result in a notable overestimation of electrical conductivity in CNT/polymer composites.

*Thermal expansion of matrix*

Lomov et al. [29] numerically estimates the factor of CTE to be less influential on CNT network's TCR than other mechanisms. Conversely, several studies have observed that the CTE of the matrix is one of the core mechanisms in determining the TCR of a CNT/polymer nanocomposite [10, 34–36]. The positive TCR likely results from the CTE of the epoxy matrix which increases the CNT-CNT electrical tunneling gap, leading to an increase in resistance [10,36].

Gong et al. [34,35] introduced a novel multiscale percolation network model that demonstrated how zero-TCR can be achieved by: (i) regulating CNT concentrations and the CTE of the matrix [34], and (ii) adjusting the competing contributions of thermally assisted tunneling transport at CNT junctions with the CTE of the matrix [35]. The two primary mechanisms are thermally assisted tunneling at CNT junctions and the CTE of the polymer matrix, which have opposing effects. As mentioned, the CTE of the matrix can widen the intertube spacing at intersecting CNT junctions, leading to a reduction in overall conductivity. Conversely, thermally assisted tunneling can improve electron transport at these junctions, thereby increasing total conductivity. Controlling CNT concentration is another useful method, particularly considering the impact of thermal expansion within the matrix is more pronounced with a sparse CNT network. Then findings indicate that, while maintaining a fixed CNT loading, selecting an appropriate polymer matrix can also effectively mitigate the influence of environmental temperature. In this context, it is important to pair lower CNT concentrations in the nanocomposite with a polymer matrix that possesses a higher CTE.

*Glass transition temperature* ($T_g$)

The glass transition temperature of a polymer matrix plays a significant role in the thermal and electrical properties of CNT/polymer nanocomposites, particularly in relation to their TCR. The negative TCR of nanocomposites above $T_g$ can be significantly improved, as the mechanism of charge transport at the CNT junctions shifts from tunneling to hopping. The experimental findings indicate that the transport in the high temperatures is dominated by hopping [37,38]. In contrast, the influence of thermal expansion of matrix results in an increased distance at CNT junctions. As the temperature increases further, there is a more rapid increase in resistance, and as the polymer matrix attain its $T_g$ the matrix expands further due to the increased mobility of the polymer chains [39], leading to enhanced positive TCR of nanocomposites above $T_g$.

Gong et al. [35] incorporated the hopping effect into their model for high temperatures. They found that at low temperatures ($T < T_g$) the thermally assisted tunneling effect is dominant, while at high temperatures ($T > T_g$) the hopping effect prevails. When temperatures are below $T_g$, tunneling and the thermal expansion of the matrix are equally important. The key difference between thermally assisted tunneling and thermally activated hopping is that tunneling occurs at excited energy levels between the bottom and top of the barrier, while hopping consistently occurs at the top of the barrier. Their findings indicate that above $T_g$, the TCR is primarily influenced by the TCR of the polymer matrix and the loading of CNTs.

*CNT agglomeration*

The performance of CNT/polymer nanocomposites is significantly influenced by the distribution of CNTs within polymers, where the CNTs have a tendency to attract one another, resulting in the formation of bundles or large clusters or agglomerates [40]. Research has demonstrated that these agglomerates can significantly reduce the electrical properties of CNT-polymer nanocomposites, particularly when contrasted with theoretical predictions that assume a homogeneous distribution of CNTs within the polymer matrix [23,29,41–43]. To achieve the theoretical performance of CNT-polymer composites, significant efforts have been undertaken to achieve a uniform dispersion of CNTs within polymer matrices [44–46]. However, even well-dispersed CNTs can re-agglomerate during the subsequent curing process [47], and this issue becomes worse at higher CNT loadings.

To achieve precise and quantitative predictions of electrical properties, researchers have developed numerical methods primarily based on the statistical percolation network theory of randomly distributed CNTs within polymers, as well as Monte Carlo simulations [48–50]. However, to date, neither analytical nor numerical approaches have successfully quantified the great decrease in electrical conduction, piezoresistivity, and piezoresistive sensitivity in CNT-polymer composites resulting from CNT agglomeration.

To resolve the gap between experiments and theory, Gong et al. [26] developed a 3D multiscale percolating network model to quantitatively evaluate the electrical conductivity and piezoresistivity of CNT-polymer composites, incorporating the effects of CNT agglomeration and deformation (including wall deformation and tube bending). Their model aligns well with experimental data [51] on CNT-polymer composites, both with and without significant agglomeration. They found that CNT agglomeration leads to decreased electrical conductivity, reduced piezoresistivity sensitivity to CNT loadings, and nonlinearity in piezoresistivity at zero strain. Similar conclusions were reached by Lebedev et al. [43].

*Wall count* (*SWCNT vs MWCNT*)

Single-walled carbon nanotubes (SWCNTs) can be classified as either metallic or semiconducting. Metallic SWCNTs show an increase in electrical resistance as temperature rises, known as positive TCR. Conversely, semiconducting SWCNTs demonstrate a decrease in resistance with increased temperature, representing a negative TCR that varies based on tube's diameter and chirality [52–54]. In multi-walled carbon nanotubes (MWCNTs), the outermost layer plays a significant role in electrical conduction, but all layers can contribute to the overall conductivity [55,56]. Generally, larger diameter MWCNTs show metallic properties with weak inter-tube coupling [57], while smaller diameter MWCNTs behave more like SWCNTs [18].

Fischer et al. [58,59], Hone et al. [60], Bae et al. [61], and Skakalova et al. [57] investigated the electrical transport phenomena in both thin films and thick mats of SWCNTs. Their findings demonstrated that both macroscopic networks of SWCNTs behave as semiconductors at low temperatures but transite to metallic behavior above room temperature, following the interrupted metallic conduction model [62]. In thin SWCNT networks, this behavior is dominated by Schottky contacts between metallic and semiconducting tubes [57,63], whereas in thicker networks, metallic tube-to-tube junctions play a more significant role [57,61,64]. In contrast, MWCNT films show semiconducting behavior across a broad temperature range, including −272 to 27 °C [65,66], −150 to 300 °C [67], −48 to 147 °C [68], 20 to 150 °C [69], and 27 to 1627 °C [70]. The conduction mechanisms in these films are dominated by defects within the tubes and inter-tube contacts at lower temperatures, while inter-tube tunneling becomes dominant at higher temperatures [66].

*Thermal residual stress*

When a two-phase material is sintered at high temperatures and then cooled down to room temperature, residual thermal stress may occur. The reason for such stress is the difference between the thermal expansion and elastic properties of the phases. Murugaraj et al. [71] presented the temperature-dependent electrical conductivity of nanocomposite films adhered to various substrates. They utilized finite element (FE) analysis to investigate potential residual stresses and their distribution within films supported by a rigid substrate, as well as the effect of these stresses on films' temperature-dependent electrical conductivity. They suggested that residual stresses affect the electronic band structure of the nanocomposites, which in turn influences charge transfer between the conducting particles. This interaction leads to variations in the hopping energies of these nanocomposite films across different substrates.

Cen-Puc et al. [72] experimentally investigated the thermoresistive response of MWCNT/polysulfone composites under heating-cooling cycles. They observed a significant difference in thermoresistive behavior between the initial cycle and following cycles, attributing this to the elimination of residual stresses/strains in the polymer, similar to a thermal annealing effect that occurs during the initial CTE of the nanocomposite [73,74]. However, changes at post-curing [1] are a complex phenomenon related not only to attenuation of residual stresses, but also to changes in resin as effecting potential barriers of tunnelling at CNT contacts.

*The degree of cure of polymer*

The extent of polymer curing plays a vital role in influencing the TCR of a CNT/polymer nanocomposite [1], as it can influence most of the aforementioned factors. Ensuring a uniform progression of the cure reaction is essential, as it significantly affects the extent of curing and the amount of residual stress. An inadequately cured polymer [1], along with the agglomeration of CNTs [26] and thermal residual stresses [10], can contribute to nonlinear TCR behavior. Non-monotonic behavior appears to be closely related to the crosslink density of the polymer matrix, since post-curing at higher temperatures minimizes the negative TCR (NTCR) to positive TCR (PTCR) transitions [10].

Jafarypouria et al. [1] investigated the effect of cure state on the TCR of a single-walled CNT/epoxy polymer nanocomposite. Their study focused on three key aspects: (1) the duration of the post-curing process, which marks the point at which the material's electrical response stabilizes; (2) the variations in TCR during different post-curing stages; and (3) the TCR of the fully post-cured material. They discovered that TCR measurements, taken before the polymer is fully cured, can be significantly affected by ongoing post-curing processes, leading to non-monotonic temperature dependence and even negative TCR values. Unbiased TCR values were observed only once the material reached a steady state, no longer influenced by heat input. Lasater et al. [75] proposed a method of in situ polymer matrix composite materials for curing process monitoring. The method of torsional braid analysis (TBA) was used in conjunction with a real-time data acquisition system to determine the viscoelastic behavior of vinyl ester (VE)/CNT/glass fiber composite. They investigated sensitivity of nanotube-based sensing to viscoelastic changes during cure. It was observed that electrical resistance was strongly related to the advancement of curing. Lee et al. [76] investigated a nanoengineered technique for in situ monitoring of cure status by employing a CNT network. Their cure status sensing experiments revealed that the electrical resistance variations of the sensor correlated with the morphological changes in the CNT network throughout the cure cycle of the composite laminates. As the polymer infiltrates the CNT network, the spacing between CNT-CNT junctions increases, resulting in a rise in resistance. Moreover, as epoxy polymer cross-linking occurs after infiltration, the chemical cure shrinkage reduces the distance between CNT-CNT junctions, leading to a reduction in resistance.

In addition, the investigations of humidity effects [77–79] show that the resistivity of CNT networks exponentially relies on the progression of water absorption. The chirality is another important factor which affect intrinsic conductivity of CNTs [18]. All mentioned factors influence the TCR of a studied material leading to complex analyses in support of experimental observations.

The objective of the present study is to achieve zero-TCR in SWCNT/epoxy nanocomposites. To this end, we experimentally investigate the TCR behavior of the samples across temperatures ranging from -15 to 80°C, with CNT loadings of 0.2 wt.%, 0.4 wt.%, 0.6 wt.%, 1.0 wt.%, and 1.2 wt.%. Five samples are examined for each CNT concentration. Before conducting the TCR tests, the fully cured state of the CNT/epoxy samples are evaluated, as this is critical to TCR behavior based on our earlier research [1]. Following this, resistance measurements are taken over a temperature range of 25-80°C, which are followed by 24 hours of post-curing and a new cycle of measurements. In total, eight measurement cycles are conducted, focusing on the CNT concentration of 0.6 wt.%.

## 2. Reported TCR Values

The mechanisms of electrical conduction become more complex when CNTs are integrated into polymers, and experimental findings in the literature are quite varied. For instance, Barrau et al. [80] investigated the DC conductivity of MWCNT/epoxy composites containing 0.4–2.5 wt% CNTs over a temperature range of 20–110 °C. These composites exhibited a NTCR, primarily influenced by the tunneling effect in the CNT network. In contrast, Alamusi et al. [36] reported a PTCR for MWCNT/epoxy nanocomposites with CNT loadings between 1 and 5 wt% over a temperature range of 60–100 °C. They attributed this behavior to the temperature-dependent tunneling effect between CNTs. Their findings indicated that the TCR increased with both temperature and MWCNT content. Cen-Puc et al. [72] illustrated a PTCR for MWCNT/polysulfone nanocomposites over a temperature range of 25–100 °C with CNT loadings ranging from 0.5 to 25 wt%. They noted that the PTCR increased as the CNT content decreased. Meanwhile, Njuguna et al. [81] observed a transition behavior from PTCR to NTCR in MWCNT/epoxy nanocomposites at concentrations of 2–3 wt% across a temperature range from -20–110 °C. Their findings represented that resistance initially increases to a local peak at around 50°C before dropping sharply to a local minimum near 80°C, subsequently followed by an increase. Utilizing differential scanning calorimetry and Raman spectroscopy analyses, they attributed this phenomenon to the physical aging of the epoxy matrix and the rearrangement of the CNT network. Additionally, MWCNT/HDPE (high-density polyethylene) nanocomposites exhibited both PTCR [82] and PTCR-to-NTCR transition responses [83,84], influenced by the CNT network and the properties of the polymer. An NTCR-to-PTCR transition behavior was also observed in the thermoresistive response of 1 wt% MWCNT/polysulfone composites within the temperature range from -110–25 °C [85]. Gong et al. [34] reported a shift from NTCR to PTCR in multi-walled carbon nanotube (MWCNT)/polymer nanocomposites created by incorporating CNTs into epoxy resin, across a temperature range of 233–383 K. They noted that at a CNT loading of approximately 3 wt%, the resistance value remained nearly constant with temperature. These differing behaviors have been documented across various matrix materials.

**Table 1:** Different TCR behaviors across various matrix materials.

| Reference | Material | TCR behavior |
|---|---|---|
| [10] | Two-component (CNT/epoxy) nanocomposites | NTCR |
| [86] | SWCNT/polycarbonate | |
| [87] | MWCNT/PEEK | |
| [88] | MWCNT/SEBS (styrene-ethylene-butylene-styrene) | |
| [89] | MWCNT/polyamide-6 | |
| [90] | MWCNT/polyurethane | |
| [29] | Aligned CNT networks (dry or impregnated with epoxy) | PTCR |
| [91] | MWCNT/polypropylene | |
| [35] | MWCNT/vinyl ester | PTCR to NTCR |
| [92] | MWCNT/PVDF (polyvinylidene fluoride) | |
| [10] | Multiscale composites (fibers with localized CNTs near the interface and epoxy) | NTCR to PTCR |
| [93] | Thin films of SWCNT/PDDA + PSS | |
| [57] | SWCNT mats | |

Jafarypouria et al. [1], for a SWCNT/epoxy nanocomposite, demonstrated that these non-monotonic TCR values appear when measurements are taken before the curing process is fully complete, as they are affected by ongoing post-curing processes. Once the epoxy achieves a fully cured state, the TCR values become monotonic, displaying a consistent increasing trend.

## 3. Core Mechanisms for Electrical Property of CNT/Polymer Composite

Due to the significant difference in electrical conductivity between the polymer matrix and carbon nanotubes (CNTs), polymer/CNT nanocomposites show percolation-like behavior, resulting in a sharp increase in electrical conductivity once the CNT content surpasses a specific threshold. This percolation phenomenon is driven by two mechanisms: (1) electron hopping (or quantum tunneling) at the nanoscale and (2) the formation of conductive networks at the microscale. In the first mechanism, electrons can hop within individual nanotubes or transfer between them, with the likelihood of this occurring predominantly influenced by the separation distance between the CNTs. At very low CNT concentrations, where this separation distance is significant, electron hopping primarily dictates the electrical conductivity of the nanocomposite. However, as the CNT content increases and the separation distance decreases, multiple adjacent CNTs may become electrically connected, resulting in the establishment of a microscale conductive network. As the concentration of CNTs continues to rise, the effect of this conductive network on electrical conductivity becomes more pronounced than that of electron hopping [95–97]. Figure 1 illustrates core mechanisms for electrical property of the CNT/polymer nanocomposite.

In our study, the primary mechanism is the formation of conductive networks at the microscale. Literature indicates that the percolation threshold for SWCNTs in polymer matrices is quite low—approximately 0.01% to 0.05%—if CNTs are well dispersed [98,99]. For our experiments, we utilized a CNT content of 0.6%, which is well above the percolation threshold, to avoid high scatter of values due to fluctuations near percolation threshold.

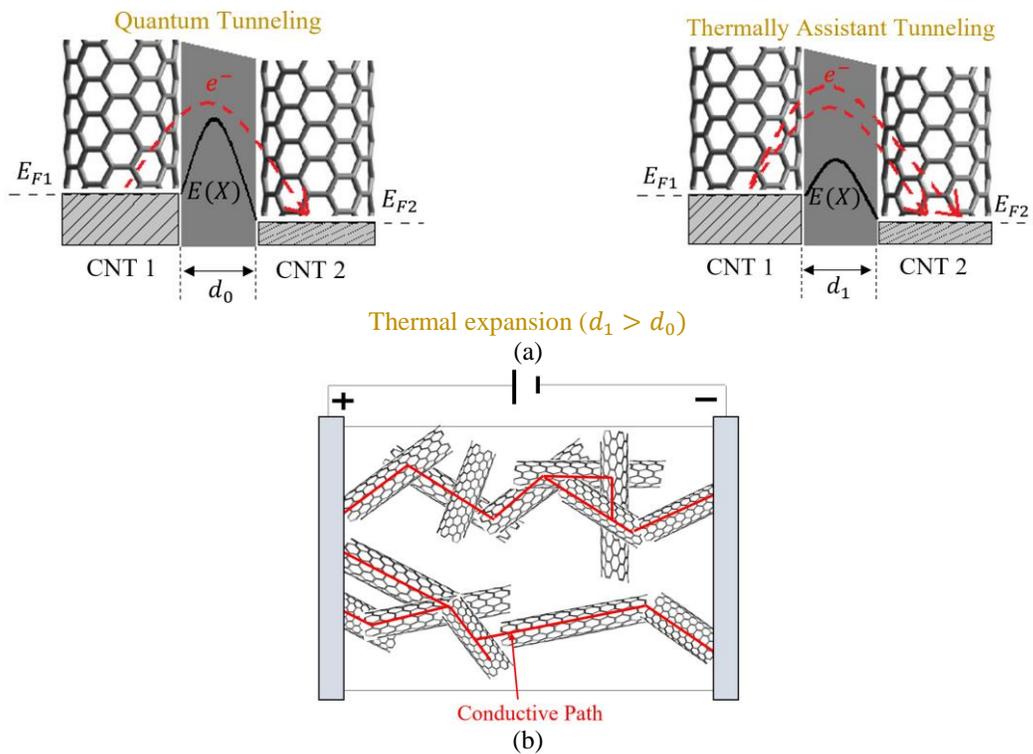

**Figure 1.** Core mechanisms for electrical property of the CNT/polymer nanocomposite. a) quantum tunneling and thermally assisted tunneling on CNT junctions. Because of thermal expansion of the polymer, the intertube space between CNTs is increased, and b) electrically conductive pathways.

## 4. Experimental

### 4.1.1 Materials

We utilized TUBALL™ MATRIX 301 (OCSiAl), a masterbatch of single-walled carbon nanotubes (SWCNTs) particularly engineered to enhance the electrical conductivity of epoxy, polyester, and polyurethane resins. Through mixing, the TUBALL™ SWCNT material is an effective filler to create conductive nanocomposites with low electrical resistance and good dispersion [98]. This is due to the good pre-dispersion of CNTs within the masterbatch, where the following shear mixing is able to dilute the mixture further with much less effort compared to the dispersion of dry CNT powder. The matrix used to create the CNT/epoxy samples was the epoxy resin system T-20-60, designed for high-performance fiber-reinforced polymer composite applications. T20-60 is a two-part low viscosity epoxy infusion resin which offers simple and flexible processing due to low viscosity and room temperature impregnation. T20-60 was developed specially for vacuum infusion molding, to produce laminates with low porosity and optimal mechanical performance.

### 4.1.2 Sample's Fabrication

The CNT/polymer nanocomposites were fabricated by mixing CNT masterbatch into epoxy resin T-20-60 to achieve the target concentrations of 0.2 wt.%, 0.4 wt.%, 0.6 wt.%, 1.0 wt.%, and 1.2 wt.% of CNTs. The resulting mixture was further mixed at a room temperature of 25°C and relative humidity of 30% with hardener T-20-60, following a resin/hardener mixing weight ratio of 100/32. As shown in Figure 2, three distinct stirring cycles were used, each with a different speed. To reduce air entrapment, a low vacuum of 0.1 mbar was used for 15 minutes in between each stirring cycle.

In order to evaluate the electrical resistance of a sample, two copper tape electrodes were placed on opposite sides of a cubic silicon mold 25 mm × 25 mm × 25 mm. Five samples, with CNT mass ratios 0.2 wt.%, 0.4 wt.%, 0.6 wt.%, 1.0 wt.%, and 1.2 wt.%, were molded and then cured at the industrial regime with temperature 130°C for 3 hours, as recommended by the resin manufacturer to achieve a high degree of cure in production. Figure 3 shows the fabricated SWCNT/epoxy samples.

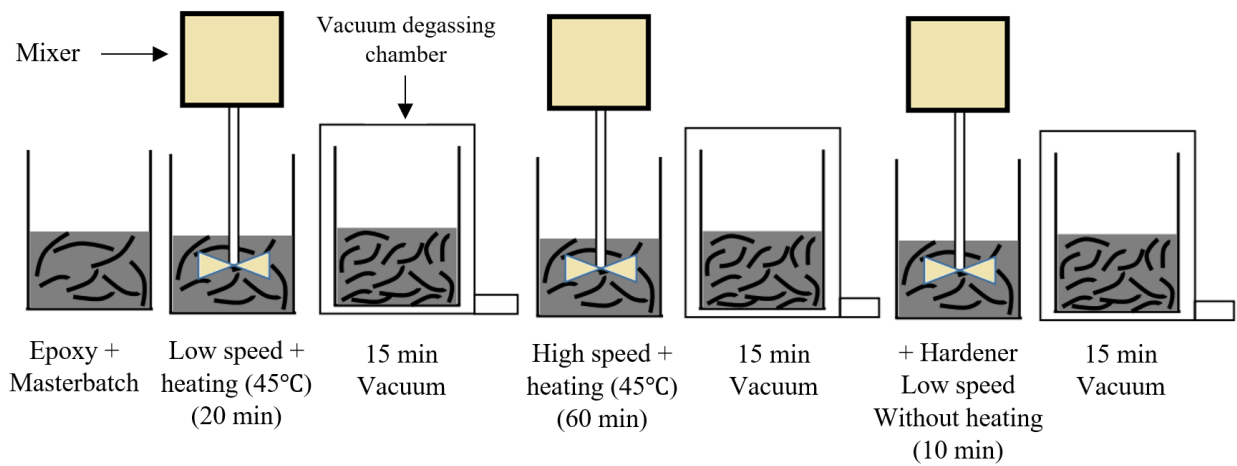

**Figure 2.** Steps involved in the synthesis process for SWCNT/epoxy samples.

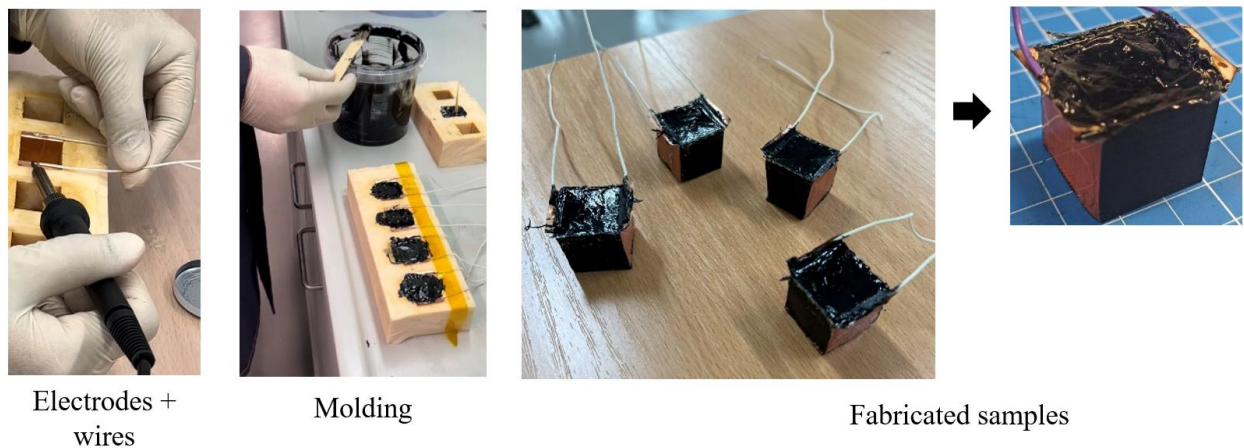

| Electrodes + wires | Molding | | Fabricated samples |

**Figure 3.** Fabrication of SWCNT/epoxy samples.

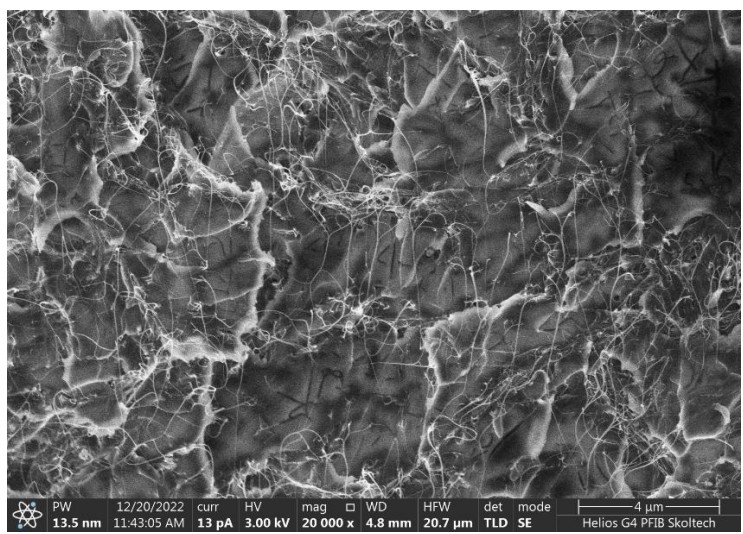

**Figure 4.** Scanning electron microscopy of a sample fracture surface.

Sample's fracture surface was examined using scanning electron microscopy to evaluate the quality of the nanofiller mixing (see Figure 4). Despite slight bundling and agglomeration, the nanofiller shows nearly ideal dispersion and distribution, displaying both homogenous and isotropic properties. This is explained by the fact that a high rotational speed of 3000 rpm resulted in a reduced resistivity and a more uniform SWCNT dispersion since it is expected [98] that electrical resistivity as well as agglomeration size and quantity decrease with increasing rotational speed.

*4.1.3 Method: Electrical Resistance Measurement*

DC electrical measurements were conducted on SWCNT/epoxy samples using a Keithley DMM6500. Copper tape electrodes were added before molding at the opposite faces of a sample, to which conducting wires were soldered (figure 3). Another end of conducting wires was connected to DMM 6500 for DC electrical measurement.

## 5. Results and Discussion

Since our earlier study [1] showed that non-monotonic TCR values appears when measurements are made before the curing process is fully complete (manufacturer recommended cure cycle e.g. [100]), the samples should first be fully cured before the TCR evaluation. Therefore, as described in section 5.1, the samples undergo a continuous post-curing procedure until they achieve epoxy T-20-60's fully cured state.

### 5.1. Fully Cured State of the Epoxy Matrix

The saturation of variations in the epoxy's electrical response (specifically, its resistance in our analysis) evaluates the duration of the post-curing process. The addition of CNTs enhances polymer's sensitivity to alterations in its molecular structure during curing. This sensitivity can be used to detect when the resin has been fully cured, as the electrical resistance will stabilize upon completion of the curing process.

This section presents resistance measurements conducted across a temperature range of 25 to 80°C, followed by the next 24 hours of post-curing and a new cycle of measurements, 8 cycles in total. The CNT concentration is 0.6% wt.%.

Since the $T_g$ of epoxy T-20-60 is 80°C, this temperature has been selected as the upper limit. When epoxy resin undergoes a curing process at temperatures exceeding its $T_g$, it shifts from a glassy state to a rubbery state, leading to loss of its fuctionality and unpredictable electrical responses during DC resistance measurements. This phenomenon also causes separation between the copper plates and the sample surface. Figure 5 illustrates the issues arising from maintaining the samples in a curing regime at 100°C for 1-2 days.

The relationship between electrical resistance and temperature for the CNT/epoxy system was investigated through eight thermal cycles, each consisting of 24 hours of post-curing at 80°C, with interruptions for measurements. The results are depicted in Figure 6. It is clear that resistance decreases significantly during post-curing until the material attains a fully cured steady state, as observed in cycles 7 and 8. Additionally, in cycle 1, the resistance exhibits a non-monotonic increase with temperature; however, in the fully cured state, this increase becomes monotonic.

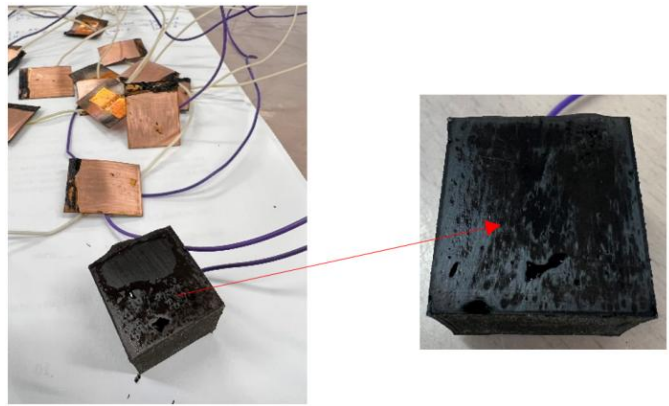

**Figure 5.** Separation of copper plates from the surface of the SWCNT/epoxy samples. Epoxy resin T-20-60 is kept at a temperature equal to 100°C (above $T_g$) for 1-2 days. The polymer changes from a glassy to a rubbery state, which results in leaking a liquid similar to oil towards the surfaces between samples and copper plates, resulting in their separation.

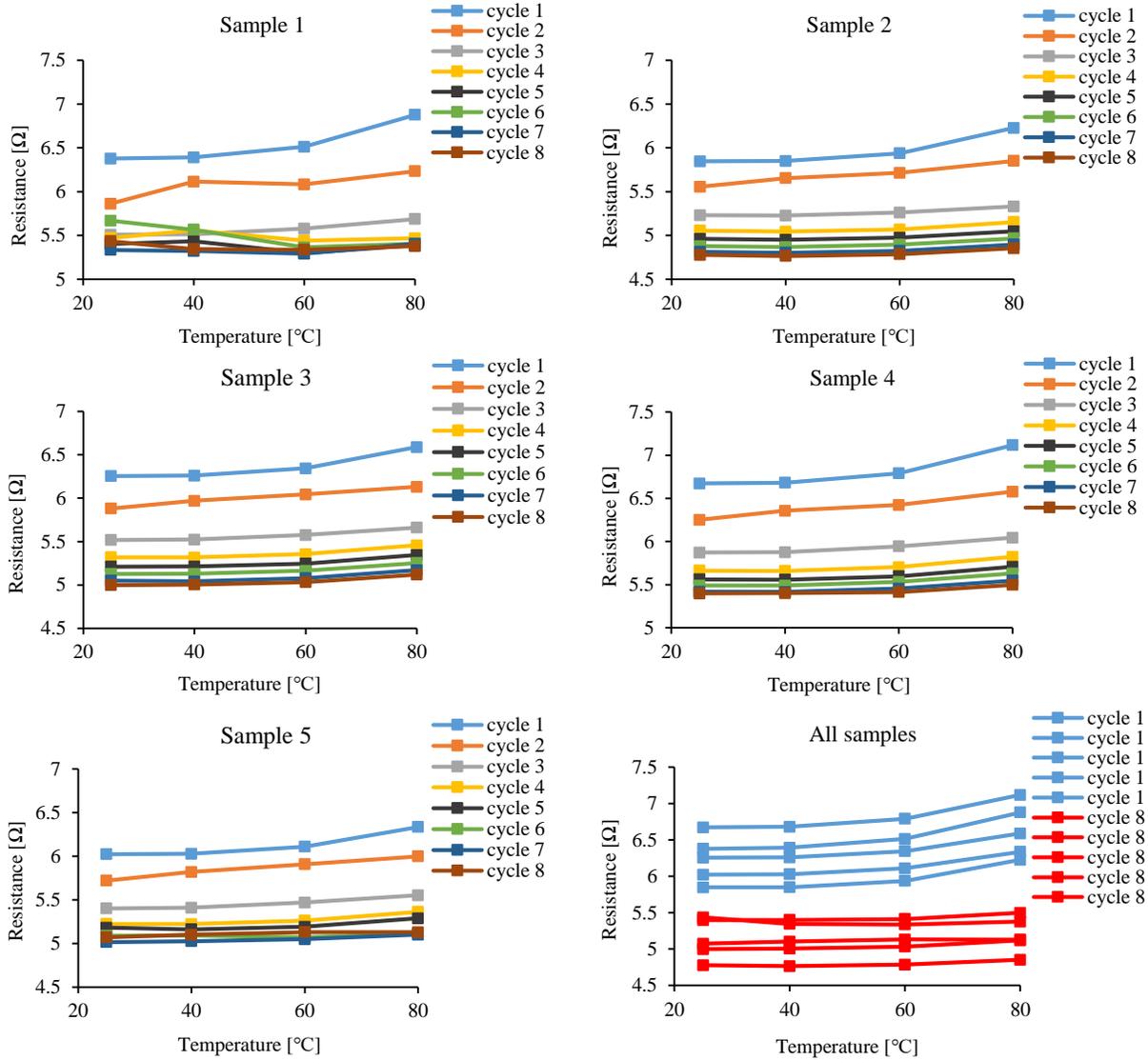

**Figure 6.** Change in electrical resistance with temperature at different post-curing cycles.

The notable reduction in resistance during the post-curing phase until achieving a fully cured state can be attributed to several mechanisms related to the curing status throughout the process [76,101,102]. Lee et al. [76] explain that the resistance variation in a CNT network is linked to the curing status, with the morphological changes in the CNT network being crucial for sensing this status.

1. **Enhanced Polymer Network:** During post-curing, the epoxy resin undergoes chemical reactions that promote greater cross-linking within the polymer network. This process decreases the distance between conductive pathways, facilitating easier electron flow.

2. **Increased Electrical Conductivity:** As curing progresses, the chemical cure shrinkage decreases the intertube distance between CNTs, resulting in a reduction in resistance.

3. **Defect Reduction:** Post-curing helps to eliminate defects and voids within the material, which can act as barriers to conductivity. A more uniform structure thereby enhances the charge transport properties of the nanocomposite.

4. **Interface Optimization:** The interactions between the CNTs and epoxy matrix may also improve during post-curing, resulting in better interface properties that further enhance electrical conduction.

## 5.2. Zero TCR Observed at Measurements

In the present section, the TCR behavior of fully cured SWCNT/epoxy nanocomposites for temperatures ranging from -15 to 80°C with CNT loadings of 0.2 wt.%, 0.4 wt.%, 0.6 wt.%, 1.0 wt.%, and 1.2 wt.% is investigated experimentally. Five samples for each CNT concentration are examined.

Figures 7a-e illustrate the TCR of SWCNT/epoxy samples for each CNT loading along with their average (Avr) of five samples. Figure 7f shows the TCRs based on Avr values. This suggests that the amount of CNT concentration is crucial in influencing the TCR characteristics of the nanocomposite. It is observed that the CNT/polymer nanocomposites show almost uniform resistance values, effectively resulting in a zero-TCR when the CNT concentration is at 1.0 wt.% (see Figure 7f). The obtained near zero-TCR for SWCNT/epoxy is attributed to: (i) finding a matching polymer matrix, (ii) controlling CNT concentration, and (iii) fully cured state of the polymer matrix. The comprehensive details regarding how these three contributions make a near zero-TCR value are discussed in introduction and section 6.

The nanocomposites with lower CNT loadings (0.2% and 0.4%) represented the highest TCR change along with non-monotonic TCR behavior for all the samples. By increasing CNT loading, non-monotonic behavior disappears and only monotonic increasing trends of TCRs are observed. This is attributed to the fact that at higher CNT loadings the main mechanism affecting the TCR behavior of CNT/polymer nanocomposites is the thermal expansion of the matrix (see section 6). Consequently, a consistent upward trend in TCR is observed as the temperature rises, since the increase in intertube distance between CNTs leads to a continues rise in the network resistance.

The non-monotonic TCR behavior at low CNT loading arises from poor network connectivity, interfacial effects, and local variations in conductivity. As the CNT loading increases, the formation of a more uniform conductive network leads to more stable and predictable resistance changes with temperature, thereby eliminating the non-monotonic behavior (see section 6). Additionally, as mentioned in the introduction, insufficiently cured polymer [1], combined with the agglomeration of CNTs [26] and thermal residual stresses [10], can lead to nonlinear TCR behavior. This non-monotonic behavior seems to have a strong correlation with the crosslink density of the polymer matrix, as post-curing at elevated temperatures reduces the transitions from negative TCR (NTCR) to positive TCR (PTCR) [10]. In this study, we aim to address these three factors contributing to nonlinear TCR behavior. A high-speed mixing for one hour, as recommended by studies conducted in our group [1,100] and other researchers [44,46,98], is performed to reduce the effects of CNT agglomeration. Also, we ensure the complete curing of the epoxy described in section 5.1 to minimize thermal residual stresses and the complexities associated with the post-curing process that hinder accurate predictions or measurements since TCR measuments with lengthy temperature equilibration serve themselves as a heat source for post-curing activation and shift in resistivity. This uncertainty in the degree of cure can influence TCR measurements, leading to biased results.

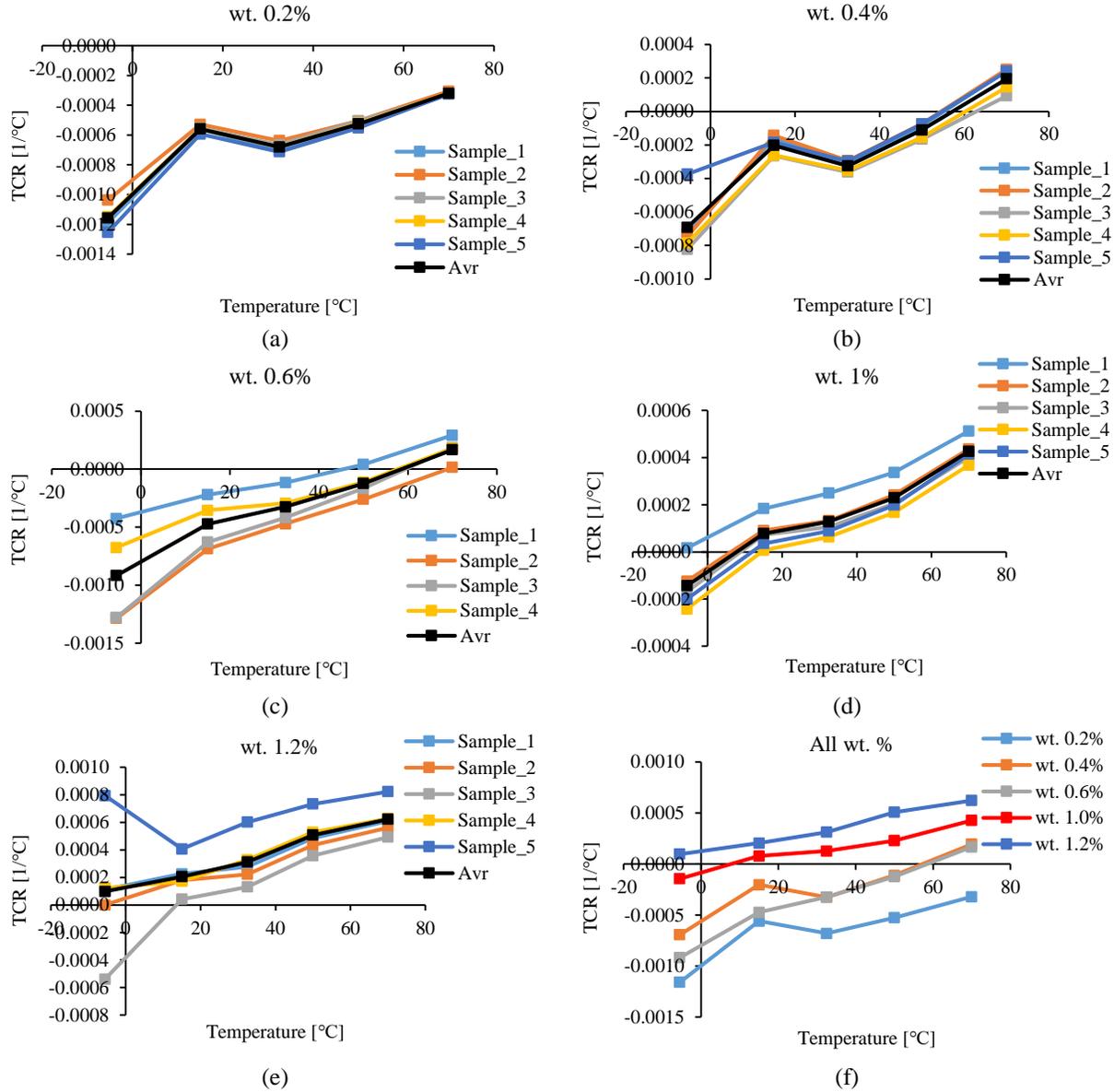

**Figure 7.** Experimental data of CNT/polymer composite TCR for temperatures from -15 to 80°C with various CNT loadings. Avr represents the average TCR values of five samples for each CNT loading.

## 6. Mechanism Analysis

As it is shown in section 3, in our study the key conductance mechanism is the formation of conductive networks at the microscale.

Contrary to the existing literature, which primarily identifies CNT intrinsic conductivity and tunneling conductance between CNTs as the main mechanisms influencing the TCR behavior of CNT/polymer nanocomposites, with the thermal expansion of the matrix considered a secondary effect, we propose a different perspective. When CNT loading reaches a level sufficient to establish a conductive pathway, the thermal expansion of the matrix becoming a dominant factor in TCR behavior. We argue that the effect is not due to the homogeneous material expansion, which has only negligible influence on tunnelling gaps, but due to configurational changes in CNT architecture because of the mismatch in coefficients of thermal expansion of polymer and CNTs. CNT's stiffness is several orders higher than that of a polymer, therefore locally (at short distances) they react to polymer movement nearly as absolutely rigid. On scales larger than the statistical Kuhn length of CNT [103], CNTs are easily bended. Therefore, polymer movements due to mismatch in coefficients of thermal expansion provide at large scale configurational changes in CNT network, easily distorted at this scale. On smaller scales, rigid CNT segments move neglecting weak polymer stiffness surrounding them. In this coupling of large scale and small scale behavior, CNT configuration can lose its stability leading to conformational changes and tunnelling gaps' rearrangements.

As the temperature increases, the expansion of the matrix results in a consistent increasing trend in TCR values. In this scenario, while electron hopping (or quantum tunneling) at the nanoscale remains a factor, it plays a reduced role in determining the overall conductivity of the network once a conductive path has been established through higher CNT loadings.

The behavior of TCR in CNT reinforced epoxy nanocomposites can exhibit complex or non-monotonic patterns at low CNT loading due to several interrelated physical effects:

1. Low CNT Loading:

- **Interfacial Effects:** At low CNT loadings, the number of CNTs might not be sufficient to form a continuous homogeneous conductive network throughout the epoxy matrix. This leads to relatively high interfacial resistance between the epoxy and the CNTs and fluctuations in resistance as temperature changes.

- **CNTs Distribution:** The CNTs in a low-loading scenario are often poorly distributed, which can cause localized areas of higher or lower conductivity. As temperature varies, the resistance measured may not correlate linearly with temperature due to these inhomogeneities.

- **Electron Scattering:** Another contributing factor is increased electron scattering at the interfaces between the CNTs and epoxy. As temperature increases, the thermal agitation affects the scattering mechanisms, which can lead to a non-monotonic TCR as some pathways become more resistant while others become less.

2. Increasing CNT Loading:

- **Formation of Conductive Networks:** As the CNT loading increases, a more continuous and interconnected network of CNTs is formed. This enhances the overall conductivity of the composite and minimizes the interfacial resistances that were significant at lower loadings, leading to more stable electrical properties.

- **Reduced Variability in Resistance:** With more CNTs present, the effects of individual discrepancies in CNT orientation and distribution become less significant, resulting in a more uniform and predictable response to temperature changes.

- **Thermal Stability:** Increased loading may also lead to better thermal conductivity within the CNT network, promoting uniform heat distribution and stabilizing the resistance with temperature, which helps eliminate the complex/non-monotonic TCR behavior observed at low loading.

In addition, type of polymer used is crucial for managing the competing influences of thermal expansion and conductance in CNT/epoxy nanocomposites, thereby facilitating the attainment of a zero TCR value. Gong et al. [34] demonstrated that, while the CNT loading is fixed, selecting a suitable polymer matrix is another effective way to reduce the influence of environment temperature. In our study, the epoxy T-20-60 has a high coefficient of thermal expansion (CTE), usually around 50 to $70 \times 10^{-6}$ /°C, which benefits a +TCR adjusting with a CNT loading prompting –TCR; consequently, a near zero-TCR is achieved.

## 7. Conclusion

In the present study, the zero-TCR behavior of SWCNT/epoxy nanocomposites was investigated experimentally for temperatures ranging from -15 to 80°C. The CNT concentrations of 0.2 wt.%, 0.4 wt.%, 0.6 wt.%, 1.0 wt.%, and 1.2 wt.% were examined, with five samples for each CNT concentration.

The CNT/polymer nanocomposites displaying nearly constant resistance values (zero-TCR) below the $T_g$ were successfully developed with a CNT loading of approximately 1 wt.%. The study indicates that key factors in achieving zero-TCR properties include the thoroughly cured state of the polymer matrix, adjusted CNT loading, and a matching choice of a polymer matrix as compensating CNT's TCR by mismatch in its and CNTs' coefficients of thermal expansion. Moreover, formation of large clusters or agglomerates of CNTs plays a significant role in the performance of CNT/polymer nanocomposites. It is also suggested that in a fully cured CNT/epoxy network with the microscale conductive pathway as the core conduction mechanism, the coefficient of thermal expansion (CTE) of the matrix is a critical characteristic, leading to a consistent increasing trend in TCR.

[40] Li, Y., Umer, R., Isakovic, A., Samad, Y. A., Zheng, L., & Liao, K. (2013). Synergistic toughening of epoxy with carbon nanotubes and graphene oxide for improved long-term performance. *RSC advances*, *3*(23), 8849-8856.

[41] Luo, D., Wang, W. X., & Takao, Y. (2007). Effects of the distribution and geometry of carbon nanotubes on the macroscopic stiffness and microscopic stresses of nanocomposites. *Composites Science and Technology*, *67*(14), 2947-2958.

[42] Ren, X., & Seidel, G. D. (2013). Computational micromechanics modeling of piezoresistivity in carbon nanotube–polymer nanocomposites. *Composite Interfaces*, *20*(9), 693-720.

[43] Lebedev, O. V., Trofimov, A., Abaimov, S. G., & Ozerin, A. N. (2019). Modeling of an effect of uniaxial deformation on electrical conductance of polypropylene-based composites filled with agglomerated nanoparticles. *International Journal of Engineering Science*, *144*, 103132.

[44] Chung, D. D. L. (2012). Carbon materials for structural self-sensing, electromagnetic shielding and thermal interfacing. *Carbon*, *50*(9), 3342-3353.

[45] Yurekli, K., Mitchell, C. A., & Krishnamoorti, R. (2004). Small-angle neutron scattering from surfactant-assisted aqueous dispersions of carbon nanotubes. *Journal of the American Chemical Society*, *126*(32), 9902-9903.

[46] Chappell, M. A., George, A. J., Dontsova, K. M., Porter, B. E., Price, C. L., Zhou, P., ... & Steevens, J. A. (2009). Surfactive stabilization of multi-walled carbon nanotube dispersions with dissolved humic substances. *Environmental Pollution*, *157*(4), 1081-1087.

[47] Ma, P. C., Mo, S. Y., Tang, B. Z., & Kim, J. K. (2010). Dispersion, interfacial interaction and re-agglomeration of functionalized carbon nanotubes in epoxy composites. *Carbon*, *48*(6), 1824-1834.

[48] Theodosiou, T. C., & Saravanos, D. A. (2010). Numerical investigation of mechanisms affecting the piezoresistive properties of CNT-doped polymers using multi-scale models. *Composites Science and Technology*, *70*(9), 1312-1320.

[49] Taya, M., Kim, W. J., & Ono, K. (1998). Piezoresistivity of a short fiber/elastomer matrix composite. *Mechanics of materials*, *28*(1-4), 53-59.

[50] Yasuoka, T., Shimamura, Y., & Todoroki, A. (2010). Electrical resistance change under strain of CNF/flexible-epoxy composite. *Advanced Composite Materials*, *19*(2), 123-138.

[51] Yin, G., Hu, N., Karube, Y., Liu, Y., Li, Y., & Fukunaga, H. (2011). A carbon nanotube/polymer strain sensor with linear and anti-symmetric piezoresistivity. *Journal of composite materials*, *45*(12), 1315-1323.

[52] Saito, R., Fujita, M., Dresselhaus, G., & Dresselhaus, U. M. (1992). Electronic structure of chiral graphene tubules. *Applied physics letters*, *60*(18), 2204-2206.

[53] Ebbesen, T. W., Lezec, H. J., Hiura, H., Bennett, J. W., Ghaemi, H. F., & Thio, T. (1996). Electrical conductivity of individual carbon nanotubes. *Nature*, *382*(6586), 54-56.

[54] Odom, T. W., Huang, J. L., Kim, P., & Lieber, C. M. (2000). Structure and electronic properties of carbon nanotubes. *The Journal of Physical Chemistry B*, *104*(13), 2794-2809.
18